\documentclass[aps,showpacs]{revtex4}
\usepackage{amssymb}
\usepackage{graphicx}
\usepackage[english]{babel}
\usepackage{indentfirst}
\usepackage{amsxtra}
\usepackage{amsmath}
\usepackage{supertabular}
\usepackage{multirow}
\usepackage [mathcal]{eucal}
\def\beq{\begin{equation}}
\def\eeq{\end{equation}}
\def\beqn{ \begin{eqnarray} }
\def\eeqn{ \end{eqnarray} }
\def\s1s2{{ \boldsymbol{\sigma}(1) \cdot \boldsymbol{\sigma}(2) }}
\def\t1t2{{ \boldsymbol{\tau}(1) \cdot \boldsymbol{\tau}(2)  }}

\newcommand{\half}{\frac{1}{2}}

\newcommand{\bq}{{\bf q}}

\newcommand{\Ls}{{\cal{L}}}
\newcommand{\eep} {$({\rm e},{\rm e}^{\,\prime}{\rm p})$ } 
\begin{document}
\noindent
\title{Mean-field calculations of exotic nuclei ground states}
\author{G. Co', V. De Donno}
\affiliation{Dipartimento di Fisica, Universit\`a del Salento, 
Lecce, Italy  and  \\ 
INFN, Sezione di Lecce, Via Arnesano, I-73100 Lecce, Italy}
\author{P. Finelli}
\affiliation{Dipartimento di Fisica Universit\`a di Bologna, Bologna,
  Italy and  
INFN, Sezione di Bologna, I-40126 Bologna, Italy}
\author{M. Grasso}
\affiliation{Insitut de Physique Nucl\'eaire, IN2P3-CNRS, Universit\'e
Paris-Sud, F-91406 Orsay Cedex, France}
\author{M. Anguiano, A. M. Lallena}
\affiliation{Departamento de F\'\i sica At\'omica, Molecular y
  Nuclear, Universidad de Granada, E-18071 Granada, Spain}
\author{C. Giusti, A. Meucci, F. D. Pacati}
\affiliation{Dipartimento di Fisica Nucleare e Teorica, 
Universit\`a di Pavia, Pavia, Italy and \\
INFN, Sezione di Pavia,  Via Bassi 6, I-27100 Pavia, Italy}
\date{\today}
\bigskip

\begin{abstract} 
We study the predictions of three mean-field theoretical approaches in
the description of the ground state properties of some spherical
nuclei far from the stability line.  We compare binding energies,
single particle spectra, density distributions, charge and neutron
radii obtained with non-relativistic Hartree-Fock calculations carried
out with both zero and finite-range interactions, and with a
relativistic Hartree approach which uses a finite-range interaction.
The agreement between the results obtained with the three different
approaches indicates that these results are more related to the basic
hypotheses of the mean-field approach rather than to its
implementation in actual calculations.
\end{abstract}

\bigskip
\bigskip
\bigskip


\maketitle

\section{Introduction}
\label{sec:intro}

The study of the properties of nuclei far from the stability valley is
one of the major topics of interest of modern nuclear physics. Wide
experimental programs of investigation are planned at nuclear
facilities now under construction \cite{riken,gsi,frib,ganil,legnaro}
and we expect that, in the next few years, a large amount of data
regarding these nuclei will be available.  From the theoretical point
of view, the main question is whether the theories and the models
which have been developed, and tested, to describe stable nuclei will
be able to perform well also in the description of these exotic
nuclei.

We classify the nuclear theories as ab-initio and effective ones.
In the first case, nucleon-nucleon interactions built
to describe observed quantities of two-body, and eventually also of
three-body, nuclear systems are employed. These theories solve the
many-body Schr\"odinger equation exactly or by making a few and well
controlled approximations. In these last years, thanks
to the advances of the computing facilities, these theories have been
applied to the description of finite nuclear systems
\cite{pie01,kam01,dea04,gan06,ari07,hag08,qua09,rot10}.  Their success
in the description of light nuclei reinforces the validity of the
basic hypotheses of the non-relativistic description of the nuclear
systems. Despite the great progress in the field, the use of these
theories for the description of medium and heavy nuclei is still
limited because of the complexity of the calculations.

Effective theories are less ambitious. Their approach to the many-body
problem is based on the mean-field (MF) assumptions.  Effects beyond
MF are taken into account in an effective manner by the
nucleon-nucleon interaction, whose parameters are chosen to reproduce
the values of some basic observable of a wide set of nuclei.
Effective theories are used to describe, and predict, nuclear
properties different from those chosen to determine the force,
as well as the properties of nuclei not included in the fit
procedure.  The calculations of observables within effective theories
are much simpler, and numerically less involved, than those of the
microscopic theories. For these reasons effective theories are widely
used in the description of medium and heavy nuclei.

Effective nuclear theories can be classified in two groups:
phenomenological and microscopic ones. We call phenomenological those
theories, and models, where the MF is globally parametrized with a
simple potential ansatz, for example a harmonic oscillator or a
Woods-Saxon well. The parameters of these wells are determined to
reproduce the experimental values of some ground state properties and,
therefore, they change for each nucleus considered.  These approaches
are extremely useful and powerful if one considers each nucleus
individually, as, for example, in the case of the Landau-Migdal theory
of finite Fermi systems \cite{spe77,kam04,gru06}.  Unfortunately, the
requirement of the experimental knowledge of some ground-state
properties of the nuclei under investigation, strongly hinders the use
of these phenomenological theories to explore experimentally unknown
regions of the nuclear chart. 

For this purpose, the microscopic MF theories \cite{ben03,vre05,sto07}
are more promising.  In these theories, the only input is the
effective nucleon-nucleon interaction, and the MF potential is
constructed by using minimization procedures which lead to the
Hartree, or to the Hartree-Fock equations when the Pauli principle is
explicitly considered. In these approaches the parameters of the
effective interaction are chosen by making a global fit of some
properties of a large set of nuclei. Since the effective interaction
is unique for all the nuclei, microscopic MF approaches are suitable
to make predictions for nuclei not yet experimentally identified.

Non relativistic Hartree-Fock (HF) calculations have a long history
and tradition in nuclear physics.  One of the approaches of major
success uses the zero-range Skyrme interaction \cite{sky56,sky58}.
Since the original formulation of Skyrme, and the seminal paper of
Vautherin and Brink \cite{vau72}, many different parametrizations of
the interaction have been proposed. Always in HF framework, a
finite-range interaction has been proposed by Gogny and collaborators
at the beginning of the 80's \cite{dec80}.  In this case HF
calculations are more involved and, consequently, also the fit
procedure. For these reasons the number of parametrizations of this
interaction available in the literature is much smaller than that of
the Skyrme interaction.

The most recent microscopic MF approach appearing in the literature
describes the nucleonic motions by using the Dirac, rather than the
Schr\"odinger, equation \cite{ser86}. 
In this relativistic approach the
nucleon-nucleon interaction is written in terms of an effective
Lagrangian which describes the exchange of some mesons whose values of
masses and coupling constants are determined to fit some global
properties of stable nuclei. Usually, in this approach only the direct
interaction matrix elements are considered, and this leads to a set of
Dirac-Hartree equations.

A short description of the three different approaches, the details of
the interactions used in the calculations, the presentation of the
observables we have investigated and of the nuclei we have considered
are given in Sec. \ref{sec:model}.  Our results are presented and
discussed in Sec. \ref{sec:res}.  Here we first discuss in
Sec. \ref{sec:inf} the results of the infinite asymmetric nuclear
matter system.  After that, we show how the three approaches behave
when they are applied to describe nuclei out of the stability valley,
in an experimentally unexplored region of the nuclear chart.  We
wanted to identify those features of the results depending on the MF
basic assumptions and disentangle them from those related to the
specific implementation of the MF model.  Specifically, we have
investigated the properties of a set of 16 nuclei, selected for their
specific characteristics. These nuclei can be grouped in four isotopic
chains, oxygen, calcium, nickel and tin; therefore our study covers a
relatively wide range of masses in the nuclear isotope chart. Since in
our calculations we assume spherical symmetry, we have selected the
spherical nuclei of each isotope chain, these are the nuclei where the
single particle (s.p.) levels are fully occupied. In
Sec. \ref{sec:ene}, we show the results concerning binding and s.p
energies of the nuclei we have considered. The most severe test on the
predictions of the three different MF models is done in
Sec. \ref{sec:distr}, where we discuss quantities directly connected
to the wave functions, i.e. density distributions and response
functions. Finally, in Sec. \ref{sec:radii}, we present some results
regarding nuclear radii and neutron skins.  After having summarized the
main results, we present in Sec. \ref{sec:con} our conclusions.

\section{The models}
\label{sec:model}

The details of the MF approaches we have adopted in our investigations
are presented in various publications, for example in Refs.
\cite{vau72,dec80,co98,typ99}, therefore we do not repeat here the
derivation of the various expressions used in our HF and Dirac-Hartree
calculations. In this section we give information about the inputs
of our calculations, essentially about the effective interactions
we have used.

We start with the older, and more exploited, approach: the HF
calculations done with the Skyrme interaction.  This interaction
contains a central, a spin-orbit and a density-dependent term and some
velocity-dependent non local components. All of them have zero range,
though the last ones can be viewed as an approximation of
finite-range.  Several parametrization have been introduced in the
past decades, many of them presented and discussed in the review of
Ref. \cite{mey03}.

In our work we have employed the SLy5 parametrization which has been
introduced by the Saclay-Lyon collaboration in the '90s
\cite{cha97,cha98a,cha98b}.  As for the SLy4 interaction, the SLy5
parametrization has been chosen to reproduce binding energies and root
mean squared (rms) radii of several nuclei, and, in addition, some
symmetric nuclear matter and neutron matter properties.  With respect
to the SLy4 force, in the fitting procedure of the SLy5 parameters the
terms of the Hamiltonian density depending on the square of the
spin-orbit density have not been neglected.  The strength of these new
terms is determined by those of the velocity-dependent terms.
Henceforth, we shall indicate as SLy5 the results obtained with this
HF approach.

Also the second MF approach we have adopted is based on
non-relativistic HF calculations, but now finite-range interactions
are used.  The motivations for using interactions with finite range in
HF calculations are well discussed in the original paper of Decharg\`e
and Gogny \cite{dec80}, where a new type of effective interaction has
been proposed. This interaction, called Gogny force, has finite range
in the traditional four central channels and two zero-range terms
which are the spin-orbit and the density dependent term.

The original parametrization of the interaction given in Ref.
\cite{dec80}, and called D1, was improved with a new one, called 
D1S \cite{ber91}, which was built to reproduce a larger set of data.
This parametrization is the most widely used, but it has the annoying
feature to produce a neutron matter equation of state whose energy per
nucleon becomes negative at large values of the neutron density. This
indicates a problem in the isospin dependent terms of the force.  To
fix this problem a large fitting project has been developed to
reproduce more than two thousand nuclear masses, hundreds of rms
charge radii, and also energy gaps \cite{cha07t,cha08}.  This
procedure has produced a new set of parameter values, the D1M force
\cite{gor09} which we have used in our calculations.  In the
following, we shall indicate as D1M the results obtained with the
finite-range HF approach.

The third approach we have used is based on the relativistic
mean-field theory, the traditional relativistic field theory where
nucleons are treated as Dirac particles moving in several classical
meson fields.  These fields describe in an average way the interaction
produced by the exchange of the corresponding mesons.  The theory
considers a defined number of mesons whose corresponding parameters,
the meson masses and the meson-nucleon coupling constants, are chosen
to reproduce empirical data.  Conventional relativistic mean-field
calculations usually include non-linear self-interaction meson
couplings, for the scalar and vector mesons, but, more recently, a new
class of relativistic functionals containing density dependent
meson-nucleon vertex functions has been introduced.  In principle, the
functional form of the meson-nucleon vertexes can be deduced from
Dirac-Brueckner calculations with realistic free-space nucleon-nucleon
interactions \cite{typ99} but, to have a better agreement with
experimental data, in this work we used a phenomenological
parametrization of the meson-nucleon coupling called DDME2
\cite{lal05}.  This parametrization has a density dependence for the
$\sigma$, $\omega$, and $\rho$ meson-nucleon couplings adjusted to
reproduce properties of symmetric nuclear matter, neutron matter, and
a limited set of spherical nuclei. More details on the Dirac-Hartree
approach can be found in Ref.  \cite{lal05}. In the following, we
shall indicate as DDME2 the results obtained with this approach.

%
\begin{table}[hb]
\begin{center}
\begin{tabular}{ccccc}
\hline\hline
&~~~~~& \multicolumn{3}{c}{Fermi level}\\
\cline{3-5}
 Nucleus && ${\rm p}$ &~~~& ${\rm n}$ \\
\hline
 $^{16}$O && $1p_{1/2}$ && $1p_{1/2}$   \\ 
 $^{22}$O && $1p_{1/2}$ && $1d_{5/2}$   \\ 
 $^{24}$O && $1p_{1/2}$ && $2s_{1/2}$   \\ 
 $^{28}$O && $1p_{1/2}$ && $1d_{3/2}$   \\ 
\hline
 $^{40}$Ca && $1d_{3/2}$ && $1d_{3/2}$   \\ 
 $^{48}$Ca && $1d_{3/2}$ && $1f_{7/2}$   \\ 
 $^{52}$Ca && $1d_{3/2}$ && $2p_{3/2}$    \\ 
 $^{60}$Ca && $1d_{3/2}$ && $2p_{1/2}$   \\ 
\hline
 $^{48}$Ni && $1f_{7/2}$ && $1d_{3/2}$   \\ 
 $^{56}$Ni && $1f_{7/2}$ && $1f_{7/2}$   \\ 
 $^{68}$Ni && $1f_{7/2}$ && $2p_{1/2}$    \\ 
 $^{78}$Ni && $1f_{7/2}$ && $1g_{9/2}$   \\ 
\hline
 $^{100}$Sn && $1g_{9/2}$ && $1g_{9/2}$   \\ 
 $^{114}$Sn && $1g_{9/2}$ && $1g_{7/2}$   \\ 
 $^{116}$Sn && $1g_{9/2}$ && $3s_{1/2}$    \\ 
 $^{132}$Sn && $1g_{9/2}$ && $1h_{11/2}$  \\ 
\hline\hline
\end{tabular}
\caption{\small Nuclei investigated in this
  work and last occupied proton and neutron s.p. 
levels, the Fermi levels, for each of them.  
}
\label{tab:nuclei}
\end{center} 
\end{table}

We conclude this section by presenting the set of isotopes we have
chosen to investigate. We list these nuclei in Table \ref{tab:nuclei}
and, for each of them, we indicate the last occupied level, the Fermi
level.  As already mentioned in the introduction, we have chosen four
isotopic chains, and within each chain we have selected those nuclei
where the number of nucleons completely fills the s.p. levels. The
deformed Hartree-Fock-Bogoliubov calculations of
Refs. \cite{hil07,del10} confirm the spherical shape of these nuclei,
therefore our approaches based on spherical symmetry are adequate to
describe them.  In these nuclei also the pairing effects are
minimized. None of the nuclei considered shows pairing effects in the
proton sector. The situation for neutrons is more complicated, since
in some of the isotopes considered the pairing is different from zero.
We have investigated the relevance of these pairing effects by
carrying on Bardeen-Cooper-Schriefer, Hartree-Fock Bogoliubov and
relativistic Hartree-Bogoliubov calculations with the three
interactions presented above. We found that the effects of the neutron
pairing on binding energies, charge and matter rms radii are at most
of a few parts in a thousand. These results induced us to neglect the
pairing in our study.

All the nuclei we have investigated resulted to be bound, except
$^{48}$Ni in DDME2 calculations, where the energy of the proton
$1f_{7/2}$ level is positive. From the experimental point of view, it
seems rather well established that the neutron drip line for the
oxygen isotopes starts with $^{26}$O \cite{sch05} and, therefore,
$^{28}$O should not be bound.

\section{Results}
\label{sec:res}

\subsection{Infinite matter}
\label{sec:inf}
The first step of our investigation consists in comparing the
  predictions of the three MF models for the equation of state (EOS)
  of infinite nuclear matter. In this study we are interested in
  comparing the different results at the saturation densities, and how
  they evolve with increasing densities.

The systems we are studying have translational invariance and constant
  nucleonic density defined as the sum $\rho=\rho_{\rm p}+\rho_{\rm
  n}$ of the proton, $\rho_{\rm p}$, and neutron, $\rho_{\rm n}$
  densities, both of them also constant.  The energy per nucleon
  $e=E/A$ for asymmetric matter is usually written as a function of
  even powers of the asymmetry parameter $\delta=(\rho_{\rm
  n}-\rho_{\rm p})/\rho$,
\beq
e(\rho,\delta)\, =\, e(\rho,0) \, + \, e_{\rm sym}(\rho) \, \delta^2 \, + \, 
{\cal{O}}(\delta^4) \, . 
\label{eq:eos}
\eeq
Around the stability minimum of symmetric nuclear matter, at density
$\rho_0$, the two coefficients of this equation are expanded in
powers of the parameter $\epsilon=(\rho-\rho_0)/(3\,\rho_0)$. For
symmetric nuclear matter we have 
\beq
e(\rho,0) \, = \, a_V \,+ \, \half K_V \, \epsilon^2  \, + \, \ldots \, ,
\eeq
where the term of first order in $\epsilon$, related to the first
derivative, is zero because 
$e(\rho,0)$ has a minimum in $\rho_0$.
In the quadratic term, related to the second derivative, the
coefficient defined as
\beq
K_V \, = \, 9 \rho_0^2 \, \left. \frac{\partial^2 e(\rho,0)}{\partial \rho^2} 
\right | _{\rho=\rho_0} 
\eeq
is called volume compression modulus. 
The second coefficient in
Eq. (\ref{eq:eos}), i.e. the symmetry energy, is expanded as
\beq
e_{\rm sym}(\rho) \, = \, a_{\rm sym} \, + \, \Ls \, \epsilon \, + \, \ldots 
\, \,.
\eeq
The coefficient 
\beq
\Ls \, =  \, 3 \rho_0 \, \left.
\frac{\partial e_{sym}(\rho)} {\partial \rho} \right | _{\rho=\rho_0} 
\label{eq:L}
\eeq
has recently attracted great attention since it is
closely related to some neutron stars properties and to the size of 
the nuclear neutron skin \cite{bro00}. 

%
\begin{table}
\begin{center}
\begin{tabular}{cccccccc}
\hline\hline
           &~~~~ & exp             & AFDMC   & CBF    & D1M    & SLy5   & DDME2 \\ 
\hline
$\rho_0$    && 0.16 $\pm$ 0.01  & 0.16   & 0.16   & 0.16   & 0.16   &  0.15 \\
$e(\rho_0,0)$ && -16.0 $\pm$ 0.1 & -16.00 & -16.00  & -16.01 & -15.98 & -16.13 \\
$K_V$         && 220 $\pm$ 30  & 276    & 269    & 217    & 228    & 278   \\ 
$e_{\rm sym}(\rho_0)$  &&  30-35          & 31.3   & 33.94  & 29.45  & 32.66  & 33.20\\ 
$\Ls$     &&  88 $\pm$ 25    & 60.10  & 58.08  & 25.41  & 48.38  & 54.74 \\ 
\hline\hline
\end{tabular}
\caption{\small Infinite nuclear matter properties 
  for various calculations. The saturation density $\rho_0 $ is
  expressed in fm$^{-3}$. All the other quantities in MeV. 
The auxiliary field diffusion Monte Carlo (AFDMC) results are from
Ref. \cite{gan10}, and those of the correlated basis function (CBF)
theory from Ref. \cite{akm98}.
}
\label{tab:matter}
\end{center} 
\end{table}

We show in Table \ref{tab:matter} the values of some nuclear matter
quantities calculated at the saturation density $\rho_0$, and we
compare them with the empirical values and with those obtained in
auxiliary field diffusion Montecarlo (AFDMC) \cite{gan10} and
correlated basis function (CBF) \cite{akm98} calculations by using 
microscopic nucleon-nucleon interactions of Argonne-Urbana type.

We observe that all the values of the saturation densities and of the
energies per nucleon agree within 2\% and 0.4\%, respectively.  The
values of $K_V$ are very similar in the two HF calculations (D1M and
SLy5), being also close to the commonly accepted empirical value. The
value obtained with DDME2 is slightly larger, but in agreement with
the result of the microscopic calculations. Also the MF $e_{\rm
sym}(\rho_0)$ are rather similar, within 6\%, while we observe large
differences in the $\Ls$ values. 

\begin{figure}[hb]
\begin{center}
\includegraphics[scale=0.4]{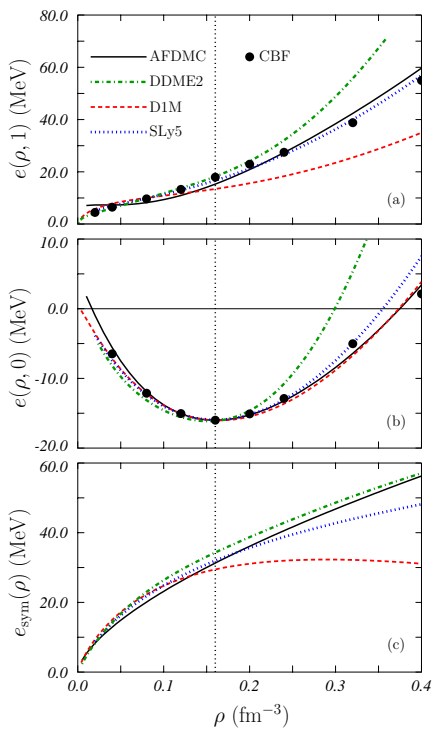} 
\caption{\small (Color on line) 
  Equation of state for pure
  neutron matter (a) and symmetric nuclear matter (b), and
  symmetry energy (c) calculated with different theories. 
  The solid circles represent
  the correlated basis function results of
  Ref. \cite{akm98}. 
  The full black lines show the auxiliary field
  diffusion Monte Carlo results of Ref. \cite{gan10}. The other
  lines show the results of the D1M (dashed red lines), SLy5 (dotted
  blue lines) and DDME2 (dashed-dotted green lines) calculations. 
  The dotted vertical lines indicate the value of
  the empirical saturation density $\rho_0$ = 0.16 fm$^{-3}$.
  }
\label{fig:eos}
\end{center}
\end{figure}

In Fig. \ref{fig:eos} we show the EOS of pure neutron matter (upper
panel), the symmetric nuclear matter (medium panel) and the symmetry
energy $e_{\rm sym}$ (lower panel).  The three MF calculations produce
very different results at large densities. The EOS generated by the
SLy5 calculations has a behavior very similar to that of the
microscopic ones. The DDME2 calculations produce stiffer EOS in both
symmetric nuclear matter and pure neutron matter. The situation for
the D1M results is more complicated. In symmetric nuclear matter there
is a good agreement with the microscopic EOS, while in pure neutron
matter the D1M EOS is the lowest one. The almost flat behavior of the
D1M symmetry energy at saturation density produces the low
value of $\Ls$ given in Table \ref{tab:matter}.

In the following section, we investigate if the differences
  we have pointed out in the nuclear matter results have consequences
  on finite nuclei observables.

\begin{figure}[ht]
\begin{center}
\includegraphics[scale=0.4]{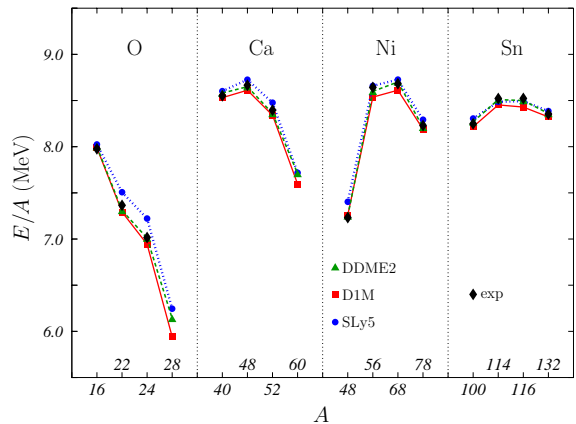} 
\caption{\small (Color on line)
  Binding energies per nucleon calculated with the three different MF
  models and compared with the experimental values \cite{aud03}.
  The lines are drawn to guide the eyes.
}
\label{fig:be}
\end{center}
\end{figure}

\subsection{Binding and single particle energies}
\label{sec:ene}

In the application of the three MF models to the description of finite
nuclear systems we first investigate binding and
s.p. energies. Binding energies are some of the observables used to
chose the values of the interaction parameters in the three
models. For this reason, we do not expect large differences between
the results obtained for this observable, even though the binding
energies of neutron rich nuclei are genuine predictions. The situation
is different for the s.p. energies which are not used to determine the
force parameters.

In Fig. \ref{fig:be} we compare the binding energies per nucleon given
by the various models with the experimental data of
Ref. \cite{aud03}. As expected, the three calculations describe
reasonably well these data.  We observe that the SLy5 results are
systematically higher than those obtained with the other
interactions. The D1M results are the most bound, and the DDME2
results are between the two.  In any case, the largest relative
difference with respect to the experimental data is 0.6\%. It is worth
pointing out that the three models produce similar binding energies
for $^{28}$O and $^{60}$Ca, that are bound in all our calculations and
for which this quantity has not been measured.

\begin{figure}[bh]
\begin{center}
\includegraphics[scale=0.4]{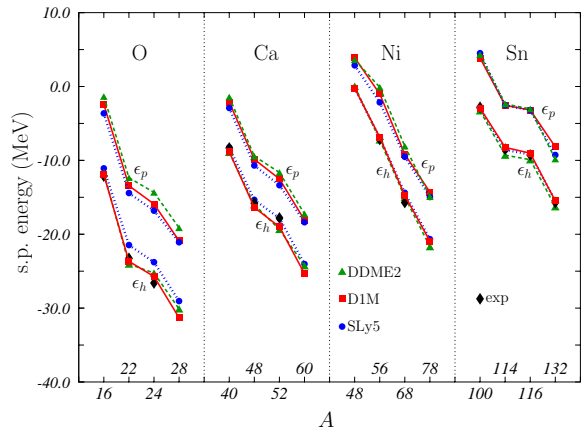} 
\caption{\small (Color on line) 
 Energies of the s.p. proton levels just below and above
 the Fermi surface,  $\epsilon_h$ and $\epsilon_p$ respectively. 
 The experimental separation energies \cite{aud03} are also shown.
}
\label{fig:spe}
\end{center}
\end{figure}

In Fig. \ref{fig:spe} we show the values of the s.p. energies of the
proton levels around the Fermi surface. Specifically, the energies,
$\epsilon_h$, of the last occupied hole level of each nucleus
indicated in Table \ref{tab:nuclei} are plotted. In our model, these
energies correspond to the proton separation energies $S_{\rm p}$
\cite{rin80}, and, in the figure, we compare them with the
experimental values \cite{aud03}. The results of all the calculations
show the same trend, in each isotope chain, the protons become more
bound as the neutron number increases.  The scale of the figure does
not show well the fact, previously remarked, that the energy of the
proton $1f_{7/2}$ level in $^{48}$Ni is slightly positive in the DDME2
calculation.  The detailed comparison between the various calculations
indicates that the SLy5 results are less bound than the other
ones. This effect is more evident in the oxygen isotopes, and
disappears in heavier nuclei. 

\begin{figure}[ht]
\begin{center}
\includegraphics[scale=0.4]{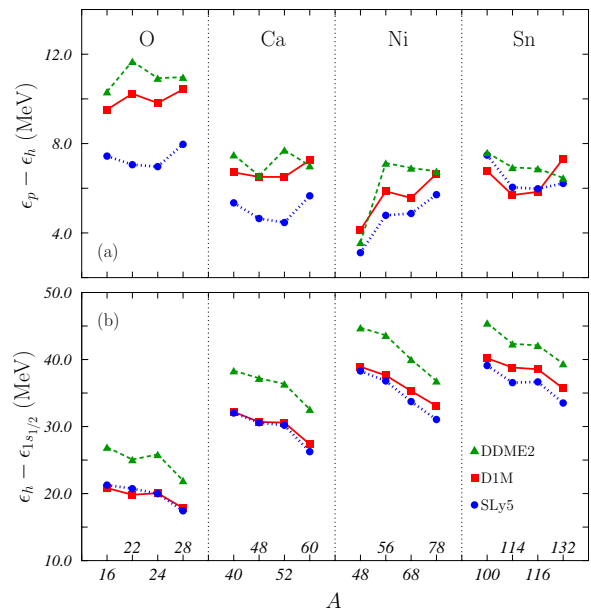} 
\caption{\small (Color on line) 
In panel (a) we show the proton energy gap,
$\epsilon_p-\epsilon_h$. In panel (b) the energy 
differences,  $\epsilon_h-\epsilon_{1s_{1/2}}$, 
between the s.p. energy of the least bound proton
level and that of the $1s_{1/2}$ level, calculated with the three
different MF models.  
}
\label{fig:hole}
\end{center}
\end{figure}

In Fig. \ref{fig:spe} we also show the s.p. energies, $\epsilon_p$, of
the proton particle levels, just above the Fermi surface. They are
$1d_{5/2}$, $1f_{7/2}$, $2p_{3/2}$ and $1g_{7/2}$ for oxygen, calcium,
nickel and tin isotopes respectively.  Also in this case we observe a
similar trend in the results of all the calculations. In each isotope
chain, the values of $\epsilon_p$ decrease with increasing neutron
number. The results obtained with the SLy5 interaction are more bound
than those obtained with the other calculations. Also in this case,
the effect is more evident in the oxygen isotopes than in heavier
nuclei.

The results shown in Fig. \ref{fig:hole} emphasize the differences 
between the three MF calculations.  In panel (a), we show the energy
gap $\epsilon_p - \epsilon_h$. In general, SLy5 produces the smallest
gaps and DDME2 the largest ones, with few exceptions. The differences
between the three calculations become smaller as nuclei become heavier
and all of them show a minimum for $^{48}$Ni.

In panel (b) of Fig. \ref{fig:hole} we show the differences between
the s.p. energy of the least, $\epsilon_h$, and most,
$\epsilon_{1s_{1/2}}$, bound proton hole levels.  As expected, this
quantity increases with the number of protons since more s.p. levels
must to be arranged in the bound system. It is interesting the fact
that, within the same isotope chain, the increase of the neutron
number reduces the value of this energy difference and, consequently,
increases the density of proton states.  The general behavior of the
three calculations is the same, but the DDME2 results are 
consistently larger than those obtained in the non-relativistic
calculations. 

\subsection{Proton, neutron and matter
distributions and electron scattering.}
\label{sec:distr}

We have seen in the previous sections that the three models produce
remarkably different results only in infinite systems at densities
much larger than those of the saturation point. The results of the
binding and s.p. energies for the isotopes we have investigated are
very similar. We want now to investigate the differences between the
wave functions generated by the three models. We have conducted this
study by calculating matter distributions, radii, and electron
scattering cross sections. These observables are more sensitive to the
details of the wave functions than the energies.

The results we have obtained for proton, neutron, and matter
distributions show the same general features in all isotope chains.
As an example of these results, we present here the distributions of
the calcium isotopes, where the effects we want to discuss are more
extreme with respect to those found in the other nuclei.

\begin{figure}[ht]
\begin{center}
\includegraphics[scale=0.4]{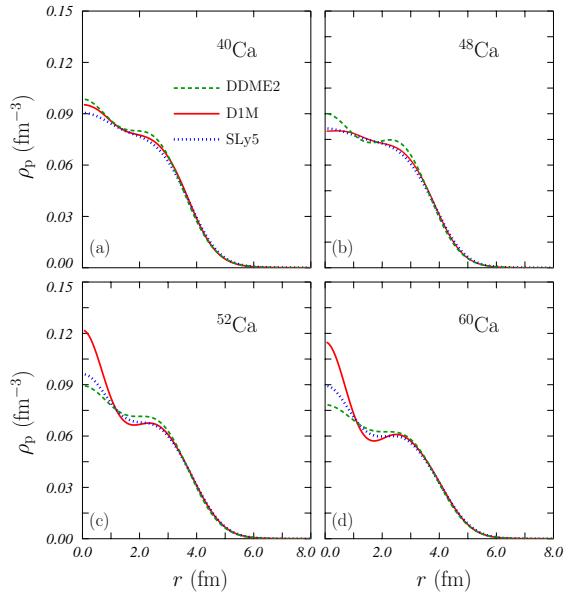} 
\caption{\small (Color on line) 
Proton distributions for the various calcium isotopes we have
considered. Full, dotted and dashed lines indicate,
respectively the D1M, Sly5 and DDME2 results.
}
\label{fig:rhopca}
\end{center}
\end{figure}

We show in Fig. \ref{fig:rhopca} the proton distributions, $\rho_{\rm
p}$, of the four calcium isotopes we have considered. There is an
excellent agreement between the results of the three calculations at
the nuclear surface.  

In the center of the nucleus, the three calculations produce similar
results for $^{40}$Ca and $^{48}$Ca, but the D1M densities are larger
than the other ones in $^{52}$Ca and $^{60}$Ca nuclei. This behavior
is a peculiarity of the D1M parametrization of the Gogny
interaction. The results obtained with the D1S force, for example, do
not show this feature \cite{don11a}. The differences with respect to
the D1M results shown in Ref. \cite{don11a} are due to an improvement
of the numerical accuracy of the HF calculations. 

\begin{figure}[b]
\begin{center}
\includegraphics[scale=0.4]{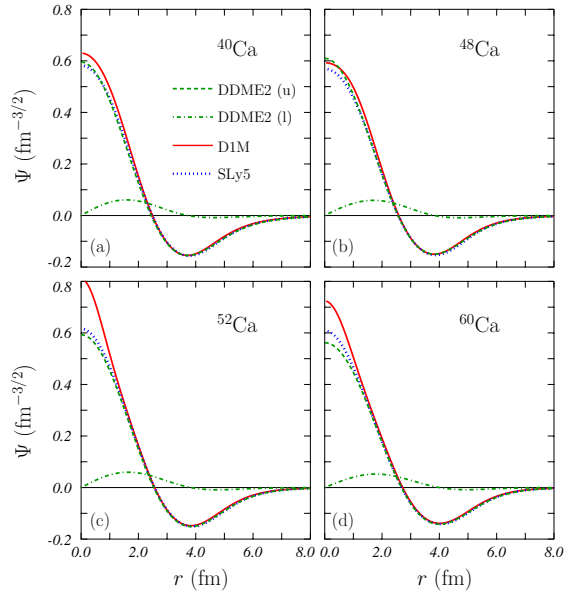} 
\caption{\small (Color on line)  Wave functions of the proton
  $2s_{1/2}$ levels for the calcium isotopes. 
  Full and dotted lines represent the D1M
  and SLy5 results, respectively. 
  For the relativistic DDME2
  calculation we show both the upper (u) and lower (l) components
  with dashed and dashed dotted lines.  }
\label{fig:2s}
\end{center}
\end{figure}

To investigate this behavior we have analyzed the $s$ wave functions,
which, in MF models are the only s.p. wave functions contributing to
density at the nuclear center. In calcium isotopes, the differences in
the density distributions are mainly due to the $2s_{1/2}$ s.p. wave
functions which we show in Fig. \ref{fig:2s}. Remarkable differences
are obtained between D1M waves and those obtained with the other MF
approaches for $^{52}$Ca and $^{60}$Ca, at $r\sim 0$ fm.  On the other
hand, it is interesting to observe the similarity of these wave
functions in the surface region. For the DDME2 calculations, we
present in the figure also the lower components of the wave functions.

A possibility of studying details of the s.p. wave functions is
offered by \eep experiments. A long series of high-precision
measurements on a wide range of nuclei
\cite{mou76,ber82,fru84,dew90,lap93,bof93,blo95,bof96} singled out
exclusive \eep knockout reactions, where the emitted proton is
measured in coincidence with the scattered electron, as the primary
tool to explore the s.p. aspects of the nuclear structure.  The
theoretical description of the \eep reaction has been developed within
the framework of the non-relativistic distorted-wave impulse
approximation (DWIA) \cite{fru84,giu87,giu88,bof93,bof96} and
relativistic distorted-wave impulse approximation (RDWIA)
\cite{pic85,udi93,udi96,kel97,udi99,
  meu01a,meu01,meu02a,meu02b,rad03}, 
including the distortion produced by the final-state interaction
between the outgoing proton and the residual nucleus, which is
described in the calculations with non-relativistic or relativistic
phenomenological optical potentials, as well as the distortion of the
electron wave functions due to the presence of the nuclear Coulomb
field. Both DWIA and RDWIA approaches were able to describe to a high
degree of accuracy \eep data on several nuclei in a wide range of
different kinematics.

\begin{figure}[hb]
\begin{center}
\includegraphics[scale=0.4]{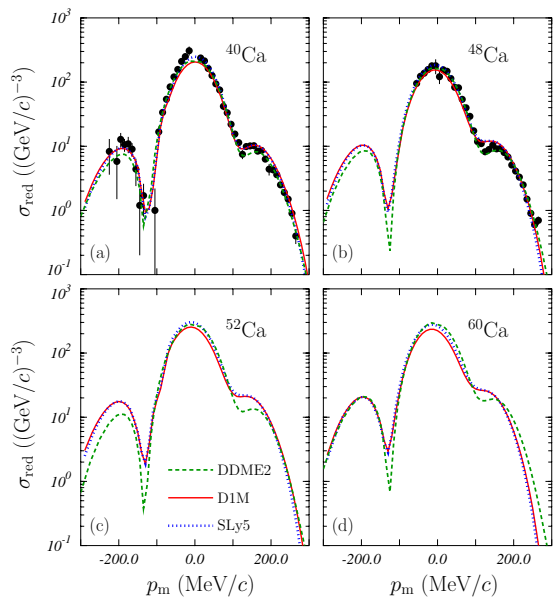} 
\caption{\small (Color on line) 
Reduced cross sections for the \eep process for various calcium 
isotopes, in parallel kinematics.
The proton is emitted from the $2s_{1/2}$ state. The
value of the energy of the incident electron is 483.2 MeV, the
electron scattering angle 61.52$^o$, the momentum transfer 450
MeV/$c$ and the energy of the emitted proton 100 MeV. The
experimental data (solid circles) of the $^{40}$Ca and $^{48}$Ca 
isotopes have been
taken from Ref. \cite{kra90t}. 
The meaning of the lines is the same as in Fig. \ref{fig:rhopca}.
}
\label{fig:eep}
\end{center}
\end{figure}

We have calculated \eep cross sections, for the emission of a proton
from the $2s_{1/2}$ s.p. level of the calcium isotopes we have
considered, under the so-called 
parallel kinematics of the NIKHEF experiments 
\cite{kra90t}. In the parallel kinematics
the momentum of the emitted proton is kept fixed and taken parallel,
or antiparallel, to the direction of the momentum transfer
$\bq$. Different values of the missing momentum $p_{\mathrm m}$, which
is the recoil momentum of the residual nucleus, are obtained by
varying the electron scattering angle and, as a consequence,
$q$. Calculations have been carried out in DWIA and RDWIA, using the
s.p. bound-state wave functions obtained in the three MF approaches
considered. Details of the calculations are described in
Ref. \cite{giu11}. 
We notice that RDWIA calculations require four-vector relativistic
wave functions for both the initial bound and the final scattering
state,as well as a relativistic nuclear current operator. 
The results of the DWIA and RDWIA calculations are compared 
in Fig. \ref{fig:eep},
where the reduced cross sections for the calcium isotopes are drawn as
a function of the missing momentum. The reduced cross section is the
cross section divided by a suitable kinematic factor~\cite{bof96} and
by the elementary electron-proton cross section~\cite{def83}.  In
order to reproduce the magnitude of the experimental data, a reduction
factor is usually applied to the calculated cross sections. For the
results shown in Fig. \ref{fig:eep} the reduction factors are 0.57 for
$^{40}$Ca and 0.58 for $^{48}$Ca. No reduction factors have been
applied for the other isotopes, where experimental data are not
available.

The differences between the various wave functions do not produce
relevant changes of the \eep cross sections in the kinematics
considered in  Fig. \ref{fig:eep}. 
We also remark the excellent agreement
between the results of all our calculations and the shape of the
experimental reduced cross sections on $^{40}$Ca and $^{48}$Ca target
nuclei. Similar results are also obtained in \cite{giu11}, where DWIA
and RDWIA calculations are performed with different s.p. bound-state
wave functions. 

\begin{figure}[ht]
\begin{center}
\includegraphics[scale=0.4]{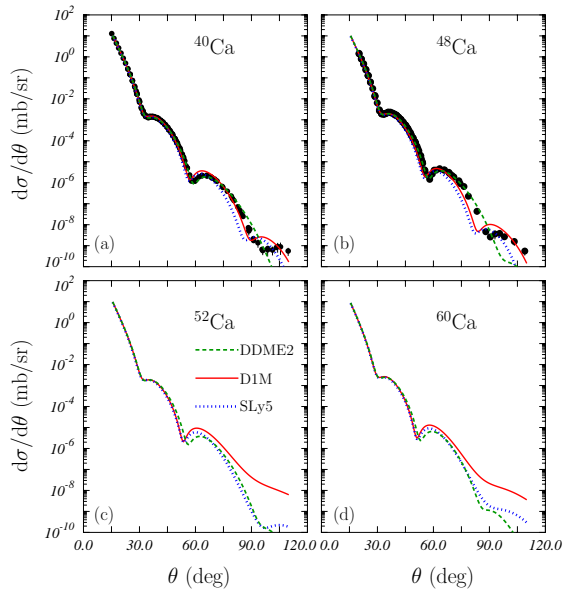} 
\caption{\small (Color on line) Elastic electron scattering cross
  sections calculated by using the charge distributions obtained with
  the three MF approaches we have considered. 
The experimental data (solid circles)
 are  from Refs. \cite{sin73,sic79,cav80t}, for $^{40}$Ca, and 
  Ref. \cite{sic75a}, for $^{48}$Ca. 
All the data have been rescaled to match an unique
  electron energy of 400 MeV. 
The meaning of the lines is the same as in Fig. \ref{fig:rhopca}.
}
\label{fig:eeca}
\end{center}
\end{figure}

Experimental information about proton distributions is obtained by the
elastic electron scattering off nuclei. In this case, the tool is
more 
sensitive to the global charge distribution of the nucleus than
to that of specific s.p. wave functions.  We have calculated elastic
electron scattering cross sections in the distorted-wave Born
approximation \cite{ann95a,ann95b} by using the proton density
distributions of Fig. \ref{fig:rhopca}. The charge densities have been
obtained by folding the proton densities with an electromagnetic
nucleon form factor of dipole form. We have verified that there are
not significant differences if other, and more sophisticated, nucleon
form factors are used.

In Fig. \ref{fig:eeca} we show the elastic electron scattering cross
sections, calculated for an electron energy of 400 MeV, as a function
of the scattering angle. We compare our calculations with the
experimental data of Refs. \cite{sin73,sic79,cav80t} for the $^{40}$Ca
nucleus and of Refs. \cite{sic75a,sic} for the $^{48}$Ca nucleus. It
is interesting to notice that, for the calcium isotopes considered, the
cross sections calculated with the different models start to differ at
about $60^o$, which, in the actual kinematics, corresponds to a
momentum transfer value of 400 MeV/$c$.

These results indicate that transferred momenta larger than 400 MeV/$c$
are necessary to produce phenomena able to disentangle the differences
between the s.p. wave functions. The \eep cross sections in the
NIKHEF kinematics \cite{kra90t} just reach this value and are not
sensitive to the differences in the $2s_{1/2}$ wave functions.

We have addressed great attention to the proton distributions because
of their connection with observables quantities, and also because they
contain information on the effective proton-neutron interaction.
Without this part of the interaction all the proton densities of an
isotope chain should be the same. We discuss now the neutron
distributions of the nuclei we have chosen to study.

\begin{figure}[ht]
\begin{center}
\includegraphics[scale=0.4]{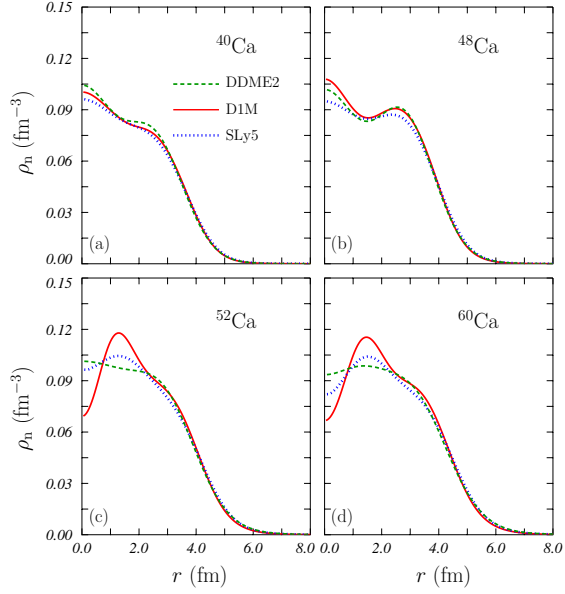} 
\caption{\small (Color on line) 
Neutron distributions for the various calcium isotopes we have considered. 
The meaning of the lines is the same as in Fig. \ref{fig:rhopca}.
}
\label{fig:rhonca}
\end{center}
\end{figure}
\begin{figure}[hb]
\begin{center}
\includegraphics[scale=0.4]{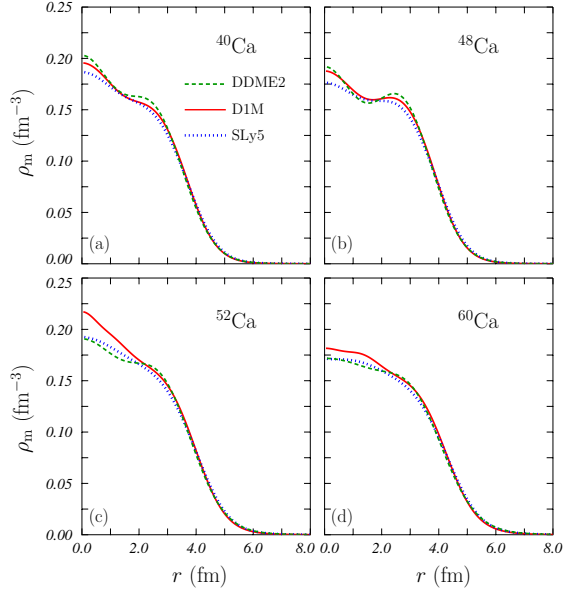} 
\caption{\small (Color on line) 
Matter distributions for the various calcium
  isotopes we have considered. 
The meaning of the lines is the same as in Fig. \ref{fig:rhopca}.
}
\label{fig:rhomca}
\end{center}
\end{figure}

In Fig. \ref{fig:rhonca} we show these distributions, $\rho_{\rm n}$,
for the calcium isotopes we have studied.  Obviously, the densities
become more extended with increasing neutron number.  Also in this
case, we observe that, for each nucleus, the densities obtained with
the different calculations have very similar surface behaviors.
As in the proton case, the differences between the various results are
remarkable in the center of the nuclei.  The major fluctuations of the
densities are shown by the D1M results and mainly in the two isotopes,
$^{52}$Ca and $^{60}$Ca, where also the proton distributions have
shown large differences with respect to the results of the other
calculations.  While in the proton case the distributions presented a
peak in the nuclear center, in the neutron case we observe a hole.

We show in Fig. \ref{fig:rhomca} the matter distributions, $\rho_{\rm
m}$, for the four calcium isotopes obtained as a sum of proton and
neutron distributions. In this case, the agreement between the results
of all the calculations is much better than that obtained by
considering separately proton and neutron distributions.  This
similarity is due to the fact that all the interactions used in the
three calculations depend on the total matter density, and not
separately on the proton or neutron densities.

The general features of the proton and neutron distributions we have
presented for the calcium isotopes are similar to those obtained in
the other isotope chains. There is excellent agreement at the surface
while differences are obtained in the nuclear interior. These
differences are larger for the separated proton and neutron densities
than in the case of the total matter density.

\begin{figure}[ht]
\begin{center}
\includegraphics[scale=0.4]{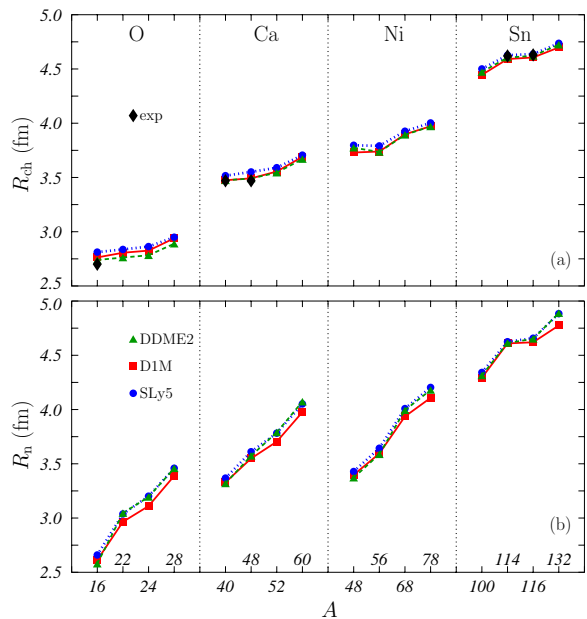} 
\caption{\small (Color on line) In the panels (a-d) we show
the rms charge radii compared with the empirical values of
Ref. \cite{ang04a}. In panels (e-f) we show the rms neutron radii. 
}
\label{fig:rms}
\end{center}
\end{figure}
\subsection{Nuclear radii and neutron skin}
\label{sec:radii}

The matter distribution results discussed in the previous sections
show differences between the various calculations in the nuclear
interior, while there is an excellent agreement at the surface. As a
consequence, the values of the rms radii calculated with the three
approaches are very similar as we show in Fig. \ref{fig:rms}. In 
panel (a) the rms charge radii are compared
with the available experimental values \cite{ang04a}. In panel (b) the
rms neutron radii are shown.

The agreement between the results of the various calculations is
remarkable.  The maximum relative differences appear between DDME2 and
SLy5 calculations in $^{24}$O for $R_{\rm ch}$ (2.8\%) and $^{16}$O
for $R_{\rm n}$ (3.0\%). The maximum absolute difference is 0.1 fm (in
$^{132}$Sn for $R_{\rm n}$).  We observe a general trend of the SLy5
calculations to produce radii slightly larger than those obtained by
the other calculations, these are, however, small differences as we
have already pointed out.

The behavior of the neutron rms radii is
rather obvious. The increasing number of neutrons increases the radius
values.  It is interesting to notice that these values are almost the
same for $^{40}$Ca and $^{48}$Ni which have same number of neutrons.
The D1M rms neutron radii are slightly smaller than the other
ones. This tendency is more evident for the Sn isotopes.

The charge radii agree very well with the available experimental
values \cite{ang04a}. We have compared our results with the values
obtained by using the semi-empirical expression given by Eq.~(4) of
Ref. \cite{pie10}, and also found excellent agreement.  In this case,
the largest difference is of about 1.9\% with the SLy5 results in the
$^{16}$O nucleus.

The behavior of the charge radii is not as obvious as that of the
neutron radii, since here the number of particles, protons, remains
the same in each isotope chain. In all the isotope chains we have
investigated, we observe a small increase with increasing neutron
number. This effect is due to the proton-neutron interaction which
rearranges the proton distributions to optimize the global matter
distribution with respect to the energy minimum.

\begin{figure}[ht]
\begin{center}
\includegraphics[scale=0.4]{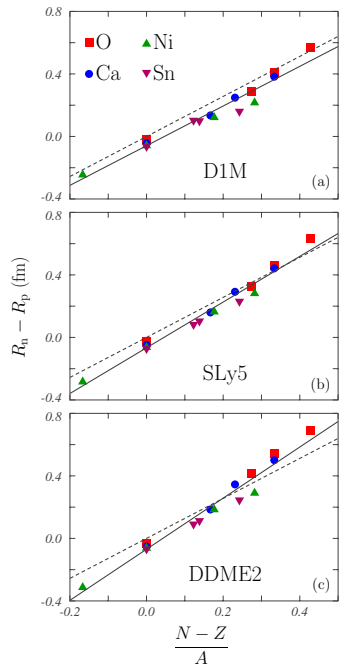} 
\caption{\small (Color on line) 
Neutron skins calculated with the three different
  approaches as a function of the relative neutron excess.  
  The full lines show a linear fit do the data and the dashed lines 
  the predictions of the model proposed in Ref. \cite{pet96}.  
}
\label{fig:skin}
\end{center}
\end{figure}

In Fig. \ref{fig:skin} we show the neutron skins, defined as
difference between neutron and proton rms radii $R_{\rm n}-R_{\rm p}$
\cite{pet96,pie10}, as a function of the relative neutron excess,
$(N-Z)/A$. The various symbols identify the different isotope chains.

We found negative values for the neutron skins of the four
self-conjugate nuclei, i.e. those with $N=Z$, and for the $^{48}$Ni
nucleus, where the number of neutrons is smaller than that of the
protons. The effect in the self-conjugate nuclei is a clear evidence
of the Coulomb repulsion, as it shown by the increase of the
phenomenon with increasing proton number. We go from -0.02 fm in
$^{16}$O to -0.08 fm in $^{100}$Sn. For these self-conjugate nuclei,
the three MF calculations produce neutron skin values which differ for
less than 0.01 fm.

In each isotope chain, the value of the neutron skins increases with
increasing neutron number, as expected. The results of the three MF
calculations slightly differ with increasing neutron number.  The
three types of calculations show an almost linear correlation between
neutron skins and the relative neutron excess.  The full lines show
the linear fits to the results of our calculations.  The three models
produce different values on the slopes of the line, specifically 1.31
fm for D1M, 1.46 fm for SLy5 and 1.63 fm for DDME2. Since for $N=Z$
the three models give essentially the same values, this indicates
that, for nuclei with neutron excess, the neutron skins obtained with
the DDME2 interaction are larger than those obtained with Sly5, and
these last ones are larger than those obtained with D1M.  The nuclear
matter properties given in table \ref{tab:matter} show a direct link
between these results and $e_{\rm sym}$ and $\Ls$, as pointed out in
the literature (see e.g. Refs. \cite{ben03,bro00,sar07,war09}).

In the figure, the dashed lines show the predictions of the model
proposed in Ref. \cite{pet96} where the neutron skins are described as
\begin{equation}
R_{\rm n}-R_{\rm p} \, = \, 1.28 \, \displaystyle \frac{N-Z}{A} \, .
\end{equation}
The model has a reasonable agreement with our linear fits, especially
with that of the D1M results.

Another interesting fact emerging from the results of
Fig. \ref{fig:skin}, is that, for similar values of the relative
neutron excess, lighter nuclei have larger neutron skins than heavier
nuclei. This is evident in the case of self-conjugate nuclei, where
the sequence O, Ca, Ni, Sn is exactly reproduced by all the
calculations.  The behavior of the neutron skins is an interesting
topic which deserves further investigation.

\section{Conclusions}
\label{sec:con}

In this work, we have compared the results of three different
implementations of the nuclear MF model in a region of the nuclear
chart which has not been yet experimentally explored. We have
investigated whether the three different MF approaches provide
different results, as it happens in infinite nuclear matter at high
density values.  We have studied the ground state properties of 16
spherical nuclei. Specifically, we have calculated binding and
s.p. energies, and also quantities more sensitive to the details of
the wave functions, such as density distributions and elastic and
inelastic electron scattering cross sections.  The general good
agreement between the results of the three models indicates that these
results are more related to the basic hypotheses of the MF model
rather than to the details of their implementation, such as the type
of interaction or the relativistic, or non-relativistic, approach.

More specifically, some of the relevant results common to all the
calculations we have presented are listed here below.
\begin{enumerate}
\item The properties of infinite nuclear matter at the saturation
  density in the three approaches are very similar and reproduce
  rather well the empirical values. The only exception is $\Ls$,
  related to the first derivative of the symmetry energy as defined in
  Eq. (\ref{eq:L}).  Above the saturation point, the behaviors of the
  EOS are remarkably different, especially in the case of pure neutron
  matter.
\item In our calculations, all the 16 nuclei investigated are
  bound. This MF prediction could be in contrast with
  the experimental evidence. We have already mentioned the fact that,
  experimentally, the neutron drip line for the oxygen isotopes starts 
  with the $^{26}$O nucleus,  
  therefore $^{28}$O is an unstable system that decays  
  by means of the strong interaction. 
\item The proton s.p. energies around the Fermi surface have similar
  values for all the three calculations. For each isotope chain
  considered, the protons are more bound when the neutron number increases. 
\item In each isotope chain, the energy available to arrange the
  proton s.p. levels decreases with increasing neutron number. As a
  consequence the density of states increases.
  We found larger density of states in non-relativistic results than
  in the relativistic ones.  
\item The study of the density distributions indicates a good
  agreement at the nuclear surface for all the three types of
  calculations. In some isotopes, we have observed very different
  behaviors in the nuclear interior when the proton and neutron
  densities are separately considered. These differences in the
  nuclear center are much smaller when the total matter distribution
  is considered.
\item 
  The large differences of the proton distributions in the nuclear
  interior are due to the $s$ proton waves. For the calcium isotopes,
  we have calculated \eep cross sections for the knock-out of a proton
  from the $2s_{1/2}$ level in the kinematics of NIKHEF experiments
  \cite{kra90t}. Despite the remarkable differences in the wave
  functions describing the $2s_{1/2}$ levels, the three calculations
  produce very similar \eep cross sections. The comparison with the
  $^{40}$Ca and $^{48}$Ca experimental data indicates that all the
  results require the same quenching factor to reproduce them.
  Calculations of elastic electron scattering cross sections show
  significant differences between the various results for momentum
  transfer values larger than 400 MeV/$c$. Since the NIKHEF
  kinematics barely reach this value, the differences between the
  various s.p. wave functions used in our calculations do not produce 
  detectable effects. 
\item The values of the rms charge radii are very similar in all the
  three calculations and agree very well with the available
  experimental data and with their empirical extrapolations.  Our
  results show a small increase of these radii with the neutron
  numbers.
\item We found an almost linear dependence of the neutron skins on the
  relative neutron excess. The relativistic calculations generated
  slightly larger skins than the other approaches.  For comparable
  values of the relative neutron excess, we found larger neutron skins
  in lighter than in heavier nuclei. This is not a trivial geometrical
  effect.
\end{enumerate}

Our investigation has been conducted for a specific set of nuclei
where deformations are absent and pairing effects negligible. In this
case, we have found a large convergence of the results of the three MF
models for all the nuclei investigated, also for those nuclei not yet
experimentally explored.  For this reason we think that producing and
investigating this type of exotic nuclei is important also from the
theoretical point of view. The comparison between the observed
properties and the MF predictions can confirm, or invalidate, the MF
model itself, and not a specific implementation of it.

\section*{ACKNOWLEDGMENTS} 
This work has been partially supported by the PRIN (Italy) {\sl
Struttura e dinamica dei nuclei fuori dalla valle di stabilit\`a}, by
the Spanish Ministerio de Ciencia e Innovaci\'on under Contract
Nos. FPA2009-14091-C02-02 and ACI2009-1007, and by the Junta de
Andaluc\'{\i}a (Grant No. FQM0220).
  
%
\newpage
%

\clearpage
\newpage
\end{document}